\documentclass{ws-mpla}
\usepackage{amsmath}
\usepackage{amssymb}
\usepackage[super]{cite}
\usepackage{bbm}
\usepackage{bm}
\usepackage{amsfonts}
\usepackage{epstopdf}
\usepackage{color}
\usepackage[colorlinks,linkcolor=red,anchorcolor=blue,citecolor=green]{hyperref}
\renewcommand{\theequation}{\thesection.\arabic{equation}}
\newcommand\diff{\mathrm{d}}

\newcommand\im{\mathrm{i}}
\newcommand\e{\mathrm{e}}
\begin{document}
\title{
EQUIVALENCE OF ZETA FUNCTION TECHNIQUE AND ABEL-PLANA FORMULA IN REGULARIZING THE CASIMIR ENERGY OF HYPER-RECTANGULAR CAVITIES }
\author{RUI-HUI LIN and XIANG-HUA ZHAI\footnote{zhaixh@shnu.edu.cn}}
\address{Shanghai United Center for Astrophysics (SUCA),\\
	Shanghai Normal University,\\
	100 Guilin Road, Shanghai 200234, China}

\maketitle

\begin{abstract}
Zeta function regularization is an effective method to extract physical significant quantities from infinite ones.
It is regarded as mathematically simple and elegant but the isolation of the physical divergency is hidden in its analytic continuation. By contrast, Abel-Plana formula method permits explicit separation of divergent terms. In regularizing the Casimir energy for a massless scalar field in a $D$-dimensional rectangular box, we give the rigorous proof of the equivalence of the two methods by deriving the reflection formula of Epstein zeta function from repeatedly application of Abel-plana formula and giving the physical interpretation of the infinite integrals.
Our study may help with the confidence of choosing any regularization method at convenience among the frequently used ones, especially the zeta function method, without the doubts of physical meanings or mathematical consistency.
\keywords{Casimir effect; zeta function regularization; Abel-Plana formula}
\end{abstract}

\ccode{PACS Nos.: 02.30.Gp, 11.10.-z}
\section{Introduction}
The extraction of physically significant quantities from ill-defined ones,
is one of the most fundamental and profound problems in quantum field theory.
The consideration of applying the Riemann zeta function as a regularization procedure can be traced back to the work of G. H. Hardy\cite{Hardy1916,Hardy1949}.
After nearly a century's development in both mathematics and physics,
ever since Dowker and Critchley\cite{Dowker1976} and Hawking\cite{Hawking1977} proposed a general zeta function method of regularization in their highly influential works,
this regularization method is now widely used in quantum physical systems.
(The readers are referred to recent works such as Refs. \refcite{Elizalde2013,Erdas2013}.)
The comprehensive review of this method has also been given\cite{Elizalde1994a,Elizalde1995,Elizalde2008,Elizalde2012}.
And the regularization of Casimir energy,
which provides a remarkable macroscopic view of quantum effect,
is a representative example of applying zeta function method.

The zeta function method, as mentioned in some of the literatures\cite{Elizalde1994,Ortenzi2004},
is considered as an elegant and unique regularization method
different from other ones such as frequency cut-off method (see e.g. Ref. \refcite{Boyer1968})
and Abel-Plana formula (see e.g. Refs. \refcite{Barton1981,Barton1982}).
The key step of this method for regularization of Casimir energy
is the analytic continuation of the corresponding zeta function.
Although in the spirit of analytic continuation
the ill-defined quantities will automatically become finite,
the isolation of the divergent part is hidden,
which gives rise to some interests. In fact, studies of heat kernel has been carried out to investigate the divergency of zeta function\cite{Kirsten2001,Tierz2001,Bordag2001,Bordag2009}.
Also, comparison between the results of this method and other methods has been addressed\cite{Ruggiero1977,Ruggiero1980,Blau1988,Elizalde1994a,Moretti1999,Bordag2001,Cavalcanti2004,Edery2006,Frank2008,Herdeiro2008,Bellucci2009,Bellucci2009a,Bordag2009,Linares2010}.
And since they mostly turn out to be in agreement,
the connection between the zeta function method and the other regularization methods has attracted some attention\cite{Svaiter1991,Svaiter1992,Beneventano1996,Kurokawa2002}. In mathematics, Butzer et al.\cite{Butzer2011} have recently proved the equivalence between any two of the summation formulae
of Euler-Maclaurin, Abel-Plana, and Poisson,
which play some part in various regularization methods,
in their finite and well-defined cases.

In comparison with zeta function regularization,
the Abel-Plana formula method has an advantage in that
it permits explicit separation of divergent terms hidden in the analytical continuation of zeta function.
The Casimir effect in rectangular boxes has been a topic for several decades
(see book \refcite{Bordag2009} and references therein and also Refs. \refcite{Lukosz1971,Mamaev1979,Mamaev1979a,Caruso1991,Hacyan1993,Gilkey1994,Actor1994,Actor1995,Li1997,Mostepanenko1997,Zheng1998,Maclay2000,Santos2000,Li2001,Inui2003,Hertzberg2005,Gusso2006,Barton2006,Jauregui2006,Hertzberg2007,Zhai2007,Lim2007,Edery2007,ZHAI2009,Rypestol2010,ZHAI2011,ZHAI2011a,FENG2014,Lin2014a,Lin2014}).
While focusing on the sign changing depending on the geometry of the configuration and the type of boundary conditions,
it is also a typical example to compare the two regularization methods
and give the interpretation of the divergent terms in vacuum energy.
In Chapter 8 of Ref. \refcite{Bordag2009}, for two and three-dimensional cases,
it is shown in detail that the finite result of the Casimir energy coincides for the two methods
and the divergent terms are properly regulated and can be interpreted as background and geometric contributions.

As indicated in the works of Schuster\cite{Schuster1999,Schuster2005} and Kurokawa and Wakayama\cite{Kurokawa2002},
which have shown that in one-dimensional case the finite part of the Abel-Plana formula results in the well-defined Riemann zeta function,
an equivalent relation between the two methods is in store in the sense of a corollary.
In this paper, by repeatedly application of Abel-Plana formula,
we demonstrate that the divergent Epstein zeta function can be expressed as the dual convergent Epstein zeta function
(this is just the reflection formula) added by a divergent integral.
Furthermore, we regulate the divergent integral by frequency cut-off method
and interpret it as background or geometric contribution depending on different boundary conditions.
With this equivalence proved, it is possible that the choice of regularization methods in Casimir effect may be made for convenience,
and comparison between ``different'' methods may not be necessary.

The paper is organized as follows.
In Sect. \ref{zetaReg}, we briefly review the calculation of the Casimir energy for massless bosonic scalar fields in a $D$-dimensional box by use of zeta function regularization.
In Sect. \ref{1d},
the equivalence between Zeta function and Abel-Plana methods in one-dimensional case is proved.
In Sect. \ref{higher},
the generalized proof of the equivalence in an arbitrary dimensional case is given.
Section \ref{conclusion} contains the conclusions and discussion.
\section{Zeta Function Regularization of the Casimir Energy}
\label{zetaReg}
\setcounter{equation}{0}
The zero point energy of a non-interacting massless scalar field in a $D$-dimensional rectangular box is given as (see e.g. Ref. \refcite{Ambjorn1983}):
\begin{equation}
	\mathcal{E}=\frac12\sum_{J}\omega_J.
	\label{ebox}
\end{equation}
For periodic boundary conditions, the modes are
\begin{equation}
	\omega_J=\sqrt{(\frac{2\pi n_1}{L_1})^2+(\frac{2\pi n_2}{L_2})^2+\cdots+(\frac{2\pi n_D}{L_D})^2},
	\label{omegaJ}
\end{equation}
and the summation over $J$ is for $n_1,n_2,\cdots,n_D$ from $-\infty$ to $+\infty$ excluding the case $n_1=n_2=\cdots=n_D=0$.
For simplicity, we take the box as a hypercube $L_1=L_2=\cdots=L_D=a$.
Then the zero point energy becomes
\begin{equation}
	\mathcal{E}(D)=\frac{\pi}a\sum_{\vec{n}\in\mathbb{Z}^D\setminus\{\vec{0}\}}(\vec{n}^2)^{\frac12},
	\label{ed}
\end{equation}
where the $D$-dimensional vector $\vec{n}$ denotes the indexes $(n_1,n_2,\cdots,n_D)$.
Obviously it is divergent.
In the zeta function regularization,
we employ the Epstein zeta function
\begin{equation}
	Z_D(s)=\sum_{\vec{n}\in\mathbb{Z}^D\setminus\{\vec{0}\}}(\vec{n}^2)^{-\frac s2}
	\label{EpsteinZeta}
\end{equation}
and the energy is then written as
\begin{equation}
	\mathcal{E}(D)=\frac\pi aZ_D(-1).
	\label{ezd}
\end{equation}
With the reflection formula of the Epstein zeta function
\begin{equation}
	\pi^{-\frac s2}\Gamma(\frac s2)Z_D(s)=\pi^{\frac{s-D}2}\Gamma(\frac{D-s}2)Z_D(D-s),
	\label{EpsteinZetaReflect}
\end{equation}
the regularized energy yields
\begin{equation}
	\mathcal{E}(D)^\text{reg.}=-\frac{\Gamma(\frac{D+1}2)}{2\pi^{\frac{D+1}2}a}Z_D(D+1).
	\label{edReg}
\end{equation}
For the case $D=1$, Epstein zeta function just degenerates to Riemann zeta function
\begin{equation}
	\zeta(s)=\sum_{n=1}^\infty \frac1{n^s},
	\label{RiemannZeta}
\end{equation}
and eq.\eqref{EpsteinZetaReflect} becomes
\begin{equation}
	\pi^{-\frac s2}\Gamma(\frac s2)\zeta(s)=\pi^{\frac{s-1}2}\Gamma(\frac{1-s}2)\zeta(1-s).
	\label{RiemannZetaReflect}
\end{equation}
Then the energy is
\[\mathcal{E}(1)=\frac12\sum_{n=-\infty}^\infty\frac{2\pi\sqrt{n^2}}{a}=\frac{2\pi}a\zeta(-1).\]
And the regularized energy is
\begin{equation}
	\quad\mathcal{E}(1)^{\text{reg.}}=-\frac1{\pi a}\zeta(2)=-\frac{\pi}{6a}.
	\label{e1}
\end{equation}
The results of eqs.\eqref{edReg} and \eqref{e1} are already finite and the regularization is done.
The reflection formulae \eqref{EpsteinZetaReflect} and \eqref{RiemannZetaReflect} are pretty much the whole story of the regularization.
\section{Equivalence to Abel-Plana Method in One-Dimensional Case}
\label{1d}
\setcounter{equation}{0}
As has been seen, the reflection formula,
which is also known as a collateral form of analytic continuation of the zeta function,
plays a key role in the regularization.
In fact, for one-dimensional, namely Riemann zeta function case,
with $s>1$, one has
\begin{equation}
\begin{split}
	\frac{\Gamma(\frac s2)}{\pi^{\frac s2}}\zeta(s)=\sum_{n=1}^\infty\frac{\Gamma(\frac s2)}{\pi^{\frac s2}}n^{-s}=&(\int_0^1+\int_1^\infty)x^{\frac s2-1}\sum_{n=1}^\infty\e^{-n^2\pi x}\diff x\\
	=&\int_0^1 x^{\frac s2-1}[\frac 1{\sqrt{x}}\sum_{n=1}^\infty\e^{-\frac{n^2\pi}x}+\frac1{2\sqrt{x}}-\frac12]\diff x\\
	&+\int_1^\infty x^{\frac s2-1}\sum_{n=1}^\infty\e^{-n^2\pi x}\diff x\\
	=&\frac1{s(s-1)}+\int_1^\infty(x^{-\frac s2-\frac12}+x^{\frac s2-1})\sum_{n=1}^\infty e^{-n^2\pi x}\diff x.
	\label{RiemannAC}
\end{split}
\end{equation}
Although the employment of the integral form of Gamma function and the Poisson summation formula
in the first and second lines requires $s>1$,
the last line is meromorphic for $s$ and remains the same with $s$ swapped for $1-s$.
So eq.\eqref{RiemannZetaReflect} is proved in this sense.
In a word,
what we actually do about the ill-defined quantity in this method of regularization
is to identify it with a finite integral form which is found in the well-defined area.
This is of course the very spirit of analytic continuation.
But basically we make an ill-defined quantity equal to a finite one.
Doubts have arised from the implicit removement of the divergency.

For this reason,
we compare it with the regularization method using Abel-Plana formula
\begin{equation}
	\sum_{n=1}^\infty u(n)=-\frac12u(0)+\int_0^\infty u(x)\diff x+\im\int_0^\infty\frac{u(\im t)-u(-\im t)}{\e^{2\pi t}-1}\diff t.
	\label{AbelPlana}
\end{equation}
The application of eq.\eqref{AbelPlana} to the one-dimensional case in Sect. \ref{zetaReg} tells that
the first term vanishes.
The second term of the energy is $\frac{2\pi}a\int_0^\infty x^{-s}\diff x$,
which is obviously divergent for $s<0$.
To illustrate the regularization of this term,
we introduce the frequency cut-off function $\exp(-\delta\frac {2\pi x}a)$,
where the parameter $\delta>0$ has to be put $\delta=0$ in the end,
and for $s=-1$ it becomes
\begin{equation}
	\frac{2\pi}a\int_0^\infty x\e^{-\delta\frac {2\pi x}a}\diff x=\frac{a}{2\pi\delta^2}.
\end{equation}
It is proportional to the ``volume'' $a$ of the one-dimensional box,
and corresponds to the vacuum energy of the free unbounded space within the volume of the box.
The physical Casimir energy should be the difference with respect to this kind of energy,
and thus this term should be subtracted.
What's left after regularization is the third term.
In Refs. \refcite{Schuster1999,Kurokawa2002,Schuster2005},
this term has been shown to be the integral form of a well-defined zeta function.
In fact, for $s<0$, one can carry out the integral
\begin{equation}
\begin{split}
	\im\int_0^\infty\frac{(\im t)^{-s}-(-\im t)^{-s}}{\e^{2\pi t}-1}\diff t=&2\sin\frac{s\pi}2\int_0^\infty\frac{t^{-s}}{\e^{2\pi t}-1}\diff t\\
	=&2\sin\frac{s\pi}2\int_0^\infty t^{-s}(\sum_{n=1}^\infty\e^{-2n\pi t})\diff t\\
	=&2\sin\frac{s\pi}2\sum_{n=1}^\infty (2\pi n)^{s-1}\Gamma(1-s).
	\label{conjugal1}
\end{split}
\end{equation}
Since $s<0$, the last summation is safe to be written as $\zeta(1-s)$, and then
\begin{equation}
	\im\int_0^\infty\frac{(\im t)^{-s}-(-\im t)^{-s}}{\e^{2\pi t}-1}\diff t=2\sin\frac{s\pi}2\Gamma(1-s)(2\pi)^{s-1}\zeta(1-s)=\frac{\pi^{s-\frac12}}{\Gamma(\frac s2)}\Gamma(\frac{1-s}2)\zeta(1-s).
	\label{conjugal2}
\end{equation}
That is, the reflection formula of Riemann zeta function is valid only after the regularization by Abel-Plana formula.
In other words,
the identification of the ill-defined quantity with the certain finite integral form
is itself a corollary,
or an equivalent form of the Abel-Plana formula regularization.
The use of the reflection relation or the analytic continuation of Riemann zeta function
implicitly removes the vacuum energy of the free unbounded space within the volume of the one-dimensional box
as the Abel-Plana formula method does explicitly.

\section{Generalization to Higher Dimensional Cases}
\label{higher}
\setcounter{equation}{0}
For a higher dimensional case,
both the mathematical analytic continuation and the equivalence
can be presented recursively utilizing the results of one-dimensional case.

In the proof the reflection relation of the Epstein zeta function \eqref{EpsteinZetaReflect},
the recurrence formula provides facilitation at length\cite{Lim2007}.
In fact, for homogeneous Epstein zeta function eq.\eqref{EpsteinZeta},
consider $Z_D(D-s)$, which is well-defined for $s<0$,
\begin{equation}
\begin{split}
	Z_D(D-s)=&Z_{D-1}(D-s)+2\sum_{\vec{n}\in\mathbb{Z}^{D-1}}\sum_{m\in\mathbb{N}}(\vec{n}^2+m^2)^{-\frac{D-s}2}\\
	=&Z_{D-1}(D-s)+\frac2{\Gamma(\frac{D-s}2)}\int_0^\infty t^{\frac{D-s}2-1}(\sum_{\vec{n}\in\mathbb{Z}^{D-1}}\sum_{m\in\mathbb{N}}\e^{-\vec{n}^2t}\e^{-m^2t})\diff t\\
	=&Z_{D-1}(D-s)+\frac{2\pi^{\frac{D-1}2}}{\Gamma(\frac{D-s}2)}\int_0^\infty t^{\frac{1-s}2-1}(\sum_{\vec{n}\in\mathbb{Z}^{D-1}}\sum_{m\in\mathbb{N}}\e^{-\frac{\pi^2\vec{n}^2}t-m^2t})\diff t\\
	=&Z_{D-1}(D-s)+\frac{2\pi^{\frac{D-1}2}}{\Gamma(\frac{D-s}2)}\zeta(1-s)\\
	&+\frac{4\pi^{\frac{D-s}2}}{\Gamma(\frac{D-s}2)}\sum_{\vec{n}\in\mathbb{Z}^{D-1}\setminus\{\vec{0}\}}\sum_{m\in\mathbb{N}}(\frac{\sqrt{\vec{n}^2}}m)^{\frac{1-s}2}K_{\frac{1-s}2}(2\pi m\sqrt{\vec{n}^2}),
	\label{CS1}
\end{split}
\end{equation}
where in the RHS of the last equal sign,
the Riemann zeta term comes from the $\vec{n}\in\{\vec{0}\}$ term of the third line,
and $K_\nu(z)$ is the modified Bessel function of the second kind resulting from the integral.
It is worth noting that
the integral form of Gamma function in the second line requires the argument $D-s>D$.
From the last equality of eq.\eqref{CS1},
recursion or mathematical induction\cite{Terras1973} will bring out the proof of eq.\eqref{EpsteinZetaReflect} easily
by use of the reflection result of Riemann zeta function and the fact that $K_\nu(z)=K_{-\nu}(z)$.
The final result of recursion
\begin{equation}
\begin{split}
	Z_D(D-s)=&\frac2{\Gamma(\frac{D-s}2)}\sum_{j=0}^{D-1}\pi^{\frac j2}\Gamma(\frac{D-s-j}2)\zeta(D-s-j)\\
	&+\frac{4\pi^{\frac{D-s}2}}{\Gamma(\frac{D-s}2)}\sum_{j=1}^{D-1}\sum_{\substack{m\in\mathbb{N}\\\vec{k}\in\mathbb{Z}^j\setminus\{\vec{0}\}}}(\frac{|\vec{k}|}m)^{\frac{D-s-j}2}K_{\frac{D-s-j}2}(2\pi m|\vec{k}|)
	\label{CS2}
\end{split}
\end{equation}
is well defined for $s<0$.
Using the analytic continuation of Riemann zeta function,
namely, eq.\eqref{RiemannZetaReflect} and eq.\eqref{RiemannAC},
one can still identify the ill-defined case of $Z_D(s),s<0$ with a finite quantity.

We still turn to the regularization using Abel-Plana formula to explore the significance of this identification.
Applying eq.\eqref{AbelPlana} in $Z_D(s)$, with $s<0$,
\begin{equation}
\begin{split}
	Z_D(s)=&\sum_{\vec{n}\in\mathbb{Z}^D\setminus\{\vec{0}\}}(\vec{n}^2)^{-\frac s2}\\=&\sum_{\vec{n}\in\mathbb{Z}^{D-1}\setminus\{\vec{0}\}}(\vec{n}^2)^{-\frac s2}+\sum_{\substack{\vec{n}\in\mathbb{Z}^{D-1}\\k\in\mathbb{Z}\setminus\{0\}}}(\vec{n}^2+k^2)^{-\frac s2}\\
	=&\sum_{\vec{n}\in\mathbb{Z}^{D-1}\setminus\{\vec{0}\}}(\vec{n}^2)^{-\frac s2}+2\sum_{\vec{n}\in\mathbb{Z}^{D-1}}\Big\{-\frac12(\vec{n}^2)^{-\frac s2}\\
	&+\int_0^\infty(\vec{n}^2+x^2)^{-\frac s2}\diff x+\im\int_0^\infty\frac{(\vec{n}^2+(\im t)^2)^{-\frac s2}-(\vec{n}^2+(-\im t)^2)^{-\frac s2}}{\e^{2\pi t}-1}\diff t\Big\}\\
	=&\Big\{\sum_{\vec{n}\in\mathbb{Z}^{D-1}\setminus\{\vec{0}\}}-\sum_{\vec{n}\in\mathbb{Z}^{D-1}}\Big\}(\vec{n}^2)^{-\frac s2}\\
	&+2\sum_{\vec{n}\in\mathbb{Z}^{D-1}}\Big\{\int_0^\infty(\vec{n}^2+x^2)^{-\frac s2}\diff x\\
	&+\im\int_0^\infty\frac{(\vec{n}^2+(\im t)^2)^{-\frac s2}-(\vec{n}^2+(-\im t)^2)^{-\frac s2}}{\e^{2\pi t}-1}\diff t\Big\}\\
	=&2\int_0^\infty(x^2)^{-\frac s2}\diff x+2\im\int_0^\infty\frac{((\im t)^2)^{-\frac s2}-((-\im t)^2)^{-\frac s2}}{\e^{2\pi t}-1}\diff t\\
	&+2\sum_{\vec{n}\in\mathbb{Z}^{D-1}\setminus\{\vec{0}\}}\int_0^\infty(\vec{n}^2+x^2)^{-\frac s2}\diff x\\
	&+2\im\sum_{\vec{n}\in\mathbb{Z}^{D-1}\setminus\{\vec{0}\}}\int_0^\infty\frac{(\vec{n}^2+(\im t)^2)^{-\frac s2}-(\vec{n}^2+(-\im t)^2)^{-\frac s2}}{\e^{2\pi t}-1}\diff t.
	\label{es4}
\end{split}
\end{equation}
In the RHS of the last equal sign,
the first term is obviously a divergent integral,
and the cancelation of it will be showed later.
The second term,
from eqs.\eqref{conjugal1} and \eqref{conjugal2}, is
\begin{equation}
	2\pi^{s-\frac12}\frac{\Gamma(\frac{1-s}2)}{\Gamma(\frac s2)}\zeta(1-s).
	\label{ies0}
\end{equation}

The last term is finite due to the exponential function in the denominator,
and with the branch point taken into account, and since $s<0<2$, it can be calculated
\begin{equation}
\begin{split}
	&\im\int_0^\infty\frac{(\vec{n}^2+(\im t)^2)^{-\frac{s}{2}}-(\vec{n}^2+(-\im t)^2)^{-\frac{s}{2}}}{\e^{2\pi t}-1}\diff t\\
	=&\im\int_0^\infty\left(\sum_{q\in\mathbb{N}}\e^{-2q\pi t}\right)(t^2-\vec{n}^2)^{-\frac s2}(\e^{-\im\frac{s\pi}2}-\e^{\im\frac{s\pi}2})\theta(t-|\vec{n}|)\\
	=&2\sin{\frac{s\pi}{2}}\sum_{q\in\mathbb{N}}\int_{|\vec{n}|}^\infty(t^2-\vec{n}^2)^{-\frac{s}{2}}\e^{-2q\pi t}\diff t\\
	=&\frac{2\pi^{\frac s2}}{\Gamma(\frac s2)}\sum_{q\in\mathbb{N}}(\frac{|\vec{n}|}{q})^{\frac{1-s}{2}}K_\frac{1-s}{2}(2q\pi|\vec{n}|),
	\label{es44}
\end{split}
\end{equation}
where in the second line, the Heaviside step function
\begin{equation}
	\theta(x)=
	\begin{cases}
		1,\quad x\ge0\\
		0,\quad x<0
	\end{cases}
	\label{step}
\end{equation}
 is introduced.

The third term of eq.\eqref{es4} is still divergent.
Similarly, with Abel-Plana formula \eqref{AbelPlana} employed on the summation over $\vec{n}$ once again, it can be written as
\begin{equation}
\begin{split}
	&2\sum_{\vec{n}\in\mathbb{Z}^{D-1}\setminus\{\vec{0}\}}\int_0^\infty(\vec{n}^2+x^2)^{-\frac s2}\diff x\\
	=&-2\int_0^\infty(x^2)^{-\frac s2}\diff x+4\int_0^\infty\diff x\int_0^\infty\diff y(x^2+y^2)^{-\frac s2}\\
	&+4\im\int_0^\infty\diff x\int_0^\infty\diff t\frac{(x^2+(\im t)^2)^{-\frac s2}-(x^2+(-\im t)^2)^{-\frac s2}}{\e^{2\pi t}-1}\\
	&+4\sum_{\vec{n}\in\mathbb{Z}^{D-2}\setminus\{\vec{0}\}}\int_0^\infty\diff x\int_0^\infty\diff y(\vec{n}^2+x^2+y^2)^{-\frac s2}\\
	&+4\im\sum_{\vec{n}\in\mathbb{Z}^{D-2}\setminus\{\vec{0}\}}\int_0^\infty\diff x\int_0^\infty\diff t\frac{(\vec{n}^2+x^2+(\im t)^2)^{-\frac s2}-(\vec{n}^2+x^2+(-\im t)^2)^{-\frac s2}}{\e^{2\pi t}-1}.
	\label{es5}
\end{split}
\end{equation}
Similar to eq.\eqref{es44}, the two finite conjugal integrals of eq.\eqref{es5} can be carried out as
\begin{equation}
\begin{split}
	&4\im\int_0^\infty\diff x\int_0^\infty\diff t\frac{(x^2+(\im t)^2)^{-\frac s2}-(x^2+(-\im t)^2)^{-\frac s2}}{\e^{2\pi t}-1}\\
	=&8\sin\frac{s\pi}2\sum_{q\in\mathbb{N}}\int_0^\infty\diff t\e^{-2q\pi t}\int_0^t\diff x(t^2-x^2)^{-\frac s2}\\
	=&\frac{2\pi^{s-1}\Gamma(1-\frac s2)}{\Gamma(\frac s2)}\zeta(2-s),
	\label{es51}
\end{split}
\end{equation}
and
\begin{equation}
\begin{split}
	&4\im\sum_{\vec{n}\in\mathbb{Z}^{D-2}\setminus\{\vec{0}\}}\int_0^\infty\diff x\int_0^\infty\diff t\frac{(\vec{n}^2+x^2+(\im t)^2)^{-\frac s2}-(\vec{n}^2+x^2+(-\im t)^2)^{-\frac s2}}{\e^{2\pi t}-1}\\
	=&8\sin\frac{s\pi}2\sum_{\vec{n}\in\mathbb{Z}^{D-2}\setminus\{\vec{0}\}}\sum_{q\in\mathbb{N}}\int_{\sqrt{\vec{n}^2}}^\infty\diff t\e^{-2q\pi t}\int_0^{\sqrt{t^2-\vec{n}^2}}\diff x(t^2-\vec{n}^2-x^2)^{-\frac s2}\\
	=&\frac{4\pi^{\frac s2}}{\Gamma(\frac s2)}\sum_{\vec{n}\in\mathbb{Z}^{D-2}\setminus\{\vec{0}\}}\sum_{q\in\mathbb{N}}(\frac{q}{|\vec{n}|})^{\frac s2-1}K_{1-\frac s2}(s\pi q|\vec{n}|)
	\label{es52}.
\end{split}
\end{equation}
Putting eqs.\eqref{es51} and \eqref{es52} back into eq.\eqref{es5} we have
\begin{equation}
\begin{split}
	&2\sum_{\vec{n}\in\mathbb{Z}^{D-1}\setminus\{\vec{0}\}}\int_0^\infty(\vec{n}^2+x^2)^{-\frac s2}\diff x\\
	=&-2\int_0^\infty(x^2)^{-\frac s2}\diff x+4\int_0^\infty\diff x\int_0^\infty\diff y(x^2+y^2)^{-\frac s2}\\
	&+4\sum_{\vec{n}\in\mathbb{Z}^{D-2}\setminus\{\vec{0}\}}\int_0^\infty\diff x\int_0^\infty\diff y(\vec{n}^2+x^2+y^2)^{-\frac s2}\\
	&+\frac{2\pi^{s-1}\Gamma(1-\frac s2)}{\Gamma(\frac s2)}\zeta(2-s)
	+\frac{4\pi^{\frac s2}}{\Gamma(\frac s2)}\sum_{\vec{n}\in\mathbb{Z}^{D-2}\setminus\{\vec{0}\}}\sum_{q\in\mathbb{N}}(\frac{q}{|\vec{n}|})^{\frac s2-1}K_{1-\frac s2}(s\pi q|\vec{n}|).
	\label{es53}
\end{split}
\end{equation}
Then substituting eqs.\eqref{ies0}, \eqref{es44} and \eqref{es53} into eq.\eqref{es4},
\begin{equation}
\begin{split}
	Z_D(s)=&4\int_0^\infty\diff x\int_0^\infty\diff y(x^2+y^2)^{-\frac s2}+2\pi^{s-\frac12}\frac{\Gamma(\frac{1-s}2)}{\Gamma(\frac s2)}\zeta(1-s)\\
	&+4\sum_{\vec{n}\in\mathbb{Z}^{D-2}\setminus\{\vec{0}\}}\int_0^\infty\diff x\int_0^\infty\diff y(\vec{n}^2+x^2+y^2)^{-\frac s2}\\
	&+\frac{2\pi^{s-1}\Gamma(1-\frac s2)}{\Gamma(\frac s2)}\zeta(2-s)
	+\frac{4\pi^{\frac s2}}{\Gamma(\frac s2)}\sum_{\vec{n}\in\mathbb{Z}^{D-2}\setminus\{\vec{0}\}}\sum_{q\in\mathbb{N}}(\frac{q}{|\vec{n}|})^{\frac s2-1}K_{1-\frac s2}(s\pi q|\vec{n}|)\\
	&+\frac{2\pi^{\frac s2}}{\Gamma(\frac s2)}\sum_{\vec{n}\in\mathbb{Z}^{D-1}\setminus\{\vec{0}\}}\sum_{q\in\mathbb{N}}(\frac{|\vec{n}|}{q})^{\frac{1-s}{2}}K_\frac{1-s}{2}(2q\pi|\vec{n}|).
	\label{es6}
\end{split}
\end{equation}

So far, no divergence has been discarded.
Using Abel-Plana formula \eqref{AbelPlana} once, we arrive at eq.\eqref{es4}.
The divergence lies in the first and the third terms.
Using Abel-Plana formula twice, we arrive at eq.\eqref{es6}.
The divergent one-dimensional integral in eq.\eqref{es4} is canceled
(this basically results from the boundary conditions,
different situation will be discussed later),
but replaced by a two-dimensional divergent integral.
The third term of eq.\eqref{es4}, the divergent summation over $\vec{n}\in\mathbb{Z}^{D-1}\setminus\{\vec{0}\}$,
splits into finite terms and another divergent summation over $\vec{n}\in\mathbb{Z}^{D-2}\setminus\{\vec{0}\}$.
One can easily see the pattern by comparing eqs.\eqref{es4} and \eqref{es6}.
Employing Abel-Plana formula on the divergent summation and repeating the procedure for another $D-2$ times,
we then have
\begin{equation}
\begin{split}
	Z_D(s)=&2^D\int_0^\infty(x_1^2+x_2^2+\cdots+x_D^2)^{-\frac s2}\diff x_1\diff x_2\cdots\diff x_D\\
	&+\frac{\pi^{s-\frac D2}\Gamma(\frac{D-s}2)}{\Gamma(\frac s2)}\Big\{\frac2{\Gamma(\frac{D-s}2)}\sum_{j=0}^{D-1}\pi^{\frac j2}\Gamma(\frac{D-s-j}2)\zeta(D-s-j)\\
	&+\frac{4\pi^{\frac{D-s}2}}{\Gamma(\frac{D-s}2)}\sum_{j=1}^{D-1}\sum_{\substack{q\in\mathbb{N}\\\vec{n}\in\mathbb{Z}^j\setminus\{\vec{0}\}}}(\frac{|\vec{n}|}q)^{\frac{D-s-j}2}K_{\frac{D-s-j}2}(2\pi q|\vec{n}|)\Big\}.
	\label{es7}
\end{split}
\end{equation}
The terms in the brace are recognized as $Z_D(D-s)$ from the eq.\eqref{CS2}.
So finally we use Abel-Plana formula to separate the divergent and convergent terms of Epstein zeta function as
\begin{equation}
\begin{split}
	Z_D(s)=&2^D\int_0^\infty(x_1^2+x_2^2+\cdots+x_D^2)^{-\frac s2}\diff x_1\diff x_2\cdots\diff x_D\\
	&+\frac{\pi^{s-\frac D2}\Gamma(\frac{D-s}2)}{\Gamma(\frac s2)}Z_D(D-s).
	\label{es8}
\end{split}
\end{equation}

In Ref.\refcite{Bordag2009}, it is shown that in two and three-dimensional cases the divergent part can be regulated and appears to be related to the geometric parameters,
i.e. perimeters, areas and volumes, of the configuration.
Similar result can also be seen here.
To see it more clearly, we consider the case that the side lengths $\{L_i,\:i=1,\cdots,D\}$ are not necessarily equal.
Taking the side lengths back in eq.\eqref{es8}, for $s=-1$, the divergent part of the energy is then
\begin{equation}
	\mathcal{E}^\text{div}=2^D\pi\int_0^\infty\sqrt{(\frac{x_1}{L_1})^2+\cdots+(\frac{x_D}{L_D})^2}\diff x_1\diff x_2\cdots\diff x_D.
	\label{d2}
\end{equation}
We still introduce the frequency cut-off function
\[\e^{-\delta\sqrt{(\frac{2\pi x_1}{L_1})^2+\cdots+(\frac{2\pi x_D}{L_D})^2}}\]
to illustrate the regularization of this term,
\begin{equation}
\begin{split}
	\mathcal{E}^\text{div}(\delta)=&2^D\pi\int_0^\infty\sqrt{(\frac{x_1}{L_1})^2+\cdots+(\frac{x_D}{L_D})^2}\e^{-\delta\sqrt{(\frac{2\pi x_1}{L_1})^2+\cdots+(\frac{2\pi x_D}{L_D})^2}}\diff^Dx\\
	=&\frac{\Gamma(1+D)(L_1L_2\cdots L_D)}{2^D\pi^{\frac D2}\delta^{1+D}\Gamma(\frac D2)},
	\label{d3}
\end{split}
\end{equation}
which is proportional to the volume of the $D$-dimensional box.
Similar to the one-dimensional case, this divergent term can be interpreted as the vacuum energy of the free unbounded space within the volume of the box.

Terms proportional to other geometric parameters will show in boundary conditions other than periodic ones.
In fact, following the procedure in Ref. \refcite{Lin2014},
the divergent part of the energy in the cases of Dirichlet and Neumann boundary conditions can be expressed as
\begin{equation}
	\mathcal{E}_{(i)}^\text{div}=(\pm\frac12)^{D-i}\frac{\pi}2\int_0^\infty\sqrt{(\frac{x_{\mu_1}}{L_{\mu_1}})^2+\cdots+(\frac{x_{\mu_i}}{L_{\mu_i}})^2}\diff^ix,
	\label{d4}
\end{equation}
where $i=1,\cdots,D$ and $\{\mu_i\}$ is a subset of $\{1,2,\cdots,D\}$,
and the signs ``$\pm$'' correspond to Neumann and Dirichlet boundary conditions, respectively.
With the frequency cut-off function introduced, these terms yield
\begin{equation}
\begin{split}
	\mathcal{E}_{(i)}^\text{div}(\delta)=&(\pm\frac12)^{D-i}\frac{\pi}2\int_0^\infty\sqrt{(\frac{x_{\mu_1}}{L_{\mu_1}})^2+\cdots+(\frac{x_{\mu_i}}{L_{\mu_i}})^2}\e^{-\delta\sqrt{(\frac{\pi x_1}{L_1})^2+\cdots+(\frac{\pi x_D}{L_D})^2}}\diff^ix\\
	=&(\pm\frac12)^{D-i}\frac{\Gamma(i+1)(L_{\mu_1}\cdots L_{\mu_i})}{2^{i}\pi^{\frac i2}\Gamma(\frac i2)\delta^{i+1}}.
	\label{d5}
\end{split}
\end{equation}
One can notice that the $i=D$ term, which is proportional to the volume of the box
and thus considered as the vacuum energy of the free unbounded space within the volume of the box,
is the same as the case of periodic boundary conditions.
The rest divergent terms,
which are obviously proportional to the other geometric parameters of the box,
are interpreted as the boundary or surface energy of the configuration.
It is easy to see that the divergent terms for Dirichlet boundary conditions in $D=2$ and $D=3$ cases coincide with the results in Ref.\refcite{Bordag2009}.
And actually, the relation of the divergent terms to the geometric parameters is consistent with the result of heat kernel.

The physical Casimir energy should be considered as the vacuum energy with these divergent terms subtracted.
In eq.\eqref{es8}, with the divergent part removed, what is left can be rearranged as
\[\pi^{-\frac s2}\Gamma(\frac s2)Z_D(s)=\pi^{\frac{s-D}2}\Gamma(\frac{D-s}2)Z_D(D-s),\]
which is exactly the reflection relation of Epstein zeta function eq.\eqref{EpsteinZetaReflect}.
In other words, the reflection relation is prescribed by the Abel-Plana formula method of regularization,
which shows the riddance of divergent terms explicitly.
Thus, we can view the analytic continuation of zeta function as implicit removement of
the vacuum energy of the unbounded space and the boundary and surface energy.

\section{Conclusions and Discussion}
\label{conclusion}
\setcounter{equation}{0}
In this paper,
we have studied the relationship between zeta function method and Abel-Plana formula method of regularization
for the Casimir effect of a massless scalar field inside a $D$-dimensional box,
in order to extend our understanding of the zeta function regularization.
The equivalence or identification of the two regularization methods has been revealed.
Using Abel-Plana formula repeatedly,
we separated the divergent Epstein zeta function into its dual convergent Epstein zeta function
and a divergent integral that can still be regulated by cut-off method.
In other words, through the demonstration of the equivalence of the two methods,
we see explicitly the geometric parameter dependent structure of the divergency hidden in the analytic continuation of zeta function.
This result is in agreement with the insight provided by the heat kernel expansion\cite{Kirsten2001,Tierz2001,Bordag2001,Bordag2009},
which is a well appreciated and effective analysis of the divergency of zeta function.

Moreover, the connection or simply the equivalence between the two methods,
together with the connection with other methods such as frequency cut-off\cite{Boyer1968,Svaiter1991,Svaiter1992,Beneventano1996,Butzer2011}
and point splitting\cite{Moretti1999},
guarantees the consistency of using ``different'' methods to regularize different parts of the Casimir energy\cite{Geyer2008,Lin2014}
or to obtain different forms of the result\cite{Edery2006,Bellucci2009,Bellucci2009a}.
So in the regularization of Casimir energy,
any of these methods can be chosen for convenience,
without worries about their physical meanings.

The equivalence between the two methods could be true for any zeta functions corresponding to specific physical configurations other than the Riemann and Epstein ones,
which is also worthy to study in the future.

\section*{Acknowledgments}
This work is partially supported by the Key Project of
Chinese Ministry of Education.(No211059), Innovation Program of
Shanghai Municipal Education Commission(11zz123) and Program of Shanghai Normal University (DXL124).

\bibliography{refs}
\bibliographystyle{ws-mpla}
\end{document}